# Cavity Quantum Electrodynamics with Anderson-localized Modes

Luca Sapienza[*], Henri Thyrrestrup, Søren Stobbe, Pedro David Garcia, Stephan Smolka, Peter Lodahl[†]

*DTU Fotonik, Department of Photonics Engineering, Technical University of Denmark, Ørsteds Plads 343, DK-2800 Kgs. Lyngby, Denmark.    http://www.fotonik.dtu.dk/quantumphotonics*

E-mail addresses: [*]lucs@fotonik.dtu.dk   [†]pelo@fotonik.dtu.dk

**A major challenge in quantum optics and quantum information technology is to enhance the interaction between single photons and single quantum emitters. Highly engineered optical cavities are generally implemented requiring nanoscale fabrication precision. We demonstrate a fundamentally different approach in which disorder is used as a resource rather than a nuisance. We generate strongly confined Anderson-localized cavity modes by deliberately adding disorder to photonic crystal waveguides. The emission rate of a semiconductor quantum dot embedded in the waveguide is enhanced by a factor of 15 on resonance with the Anderson-localized mode and 94 % of the emitted single-photons couple to the mode. Disordered photonic media thus provide an efficient platform for quantum electrodynamics offering an approach to inherently disorder-robust quantum information devices.**

The interaction between a single photon and a single quantized emitter is the core of cavity quantum electrodynamics (QED) and constitutes a node in a quantum information network (*1,2*). So far, cavity QED experiments have been realized with a wide range of two-level systems including atoms (*3*), ions (*4*), Cooper-pair boxes (*5*), and semiconductor quantum dots (*6-8*) coupled to photons confined in a cavity.

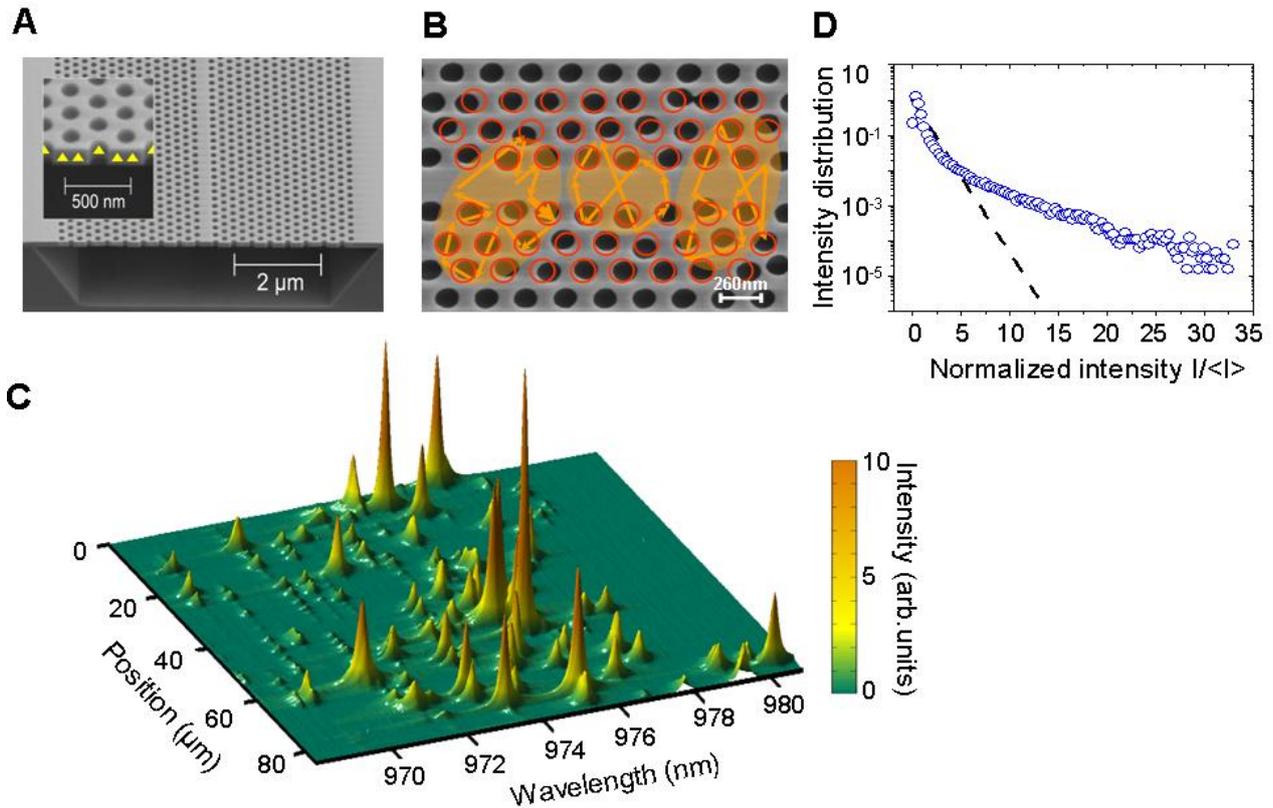

**Fig.1:** Anderson localization in disordered photonic crystal waveguides. **(A)** Scanning electron micrograph of a photonic crystal membrane waveguide without engineered disorder, containing a single layer of QDs (yellow symbols, not to scale). **(B)** Photonic crystal waveguide with 6% engineered disorder. The red circles represent the hole positions in an ideal structure without disorder. The orange arrows depict the wavevectors of localized modes. **(C)** High power photoluminescence spectra collected while scanning the excitation and collection microscope objective along the waveguide at 10 K for a 3% disordered sample. **(D)** Measured probability distribution of the normalized photoluminescence intensity extracted from the data presented in panel (C). The black dashed line represents the Rayleigh distribution.

A common requirement for all these implementations is highly engineered cavities, in some cases requiring nanometer scale accuracy (*9*). Surprisingly, multiple scattering of photons in disordered dielectric structures offers an alternative route to light confinement. If the scattering is very pronounced, Anderson-localized modes form spontaneously. Anderson localization (*10*) is a multiple scattering wave phenomenon that has been observed for, e.g., light (*11*), acoustic waves (*12*), and atomic Bose-Einstein condensates (*13*). We demonstrate cavity QED with Anderson-localized modes by efficiently coupling a single quantum dot (QD) to a disorder-induced cavity mode (*14*) in a photonic crystal waveguide.

Photonic crystals are composite nanostructures where a periodic modulation of the refractive index forms a photonic band gap of frequencies where light propagation is fully suppressed. By deliberately introducing a missing row of holes in a two-dimensional photonic crystal membrane, the periodicity is broken locally and light is guided (Fig. 1A). Such photonic crystal waveguides are strongly dispersive, i.e. light propagation depends sensitively on the optical frequency and can be slowed down. Engineering the photonic crystal waveguide enables enhancing light-matter interaction, which is required for high-efficiency single-photon sources (*15*) for quantum information technology (*1,16*). In the slow-light regime of photonic crystal waveguides, light propagation is very sensitive to unavoidable structural imperfections (*17,18*) and multiple scattering events randomize propagation (*19*). As opposed to the common belief that multiple scattering is a nuisance for a device leading to losses, here the influence of wave interference in multiple scattering stops light propagation and forms strongly confined Anderson-localized modes (*10*) (Fig. 1B). Anderson-localized modes in a photonic crystal waveguide appear due to the primarily one-dimensional nature of the propagation of light provided that the localization length is shorter than the length of the waveguide (*20*).

We deliberately create Anderson-localized modes by fabricating photonic crystal waveguides with a lithographically controlled amount of disorder (Fig. 1B). The hole positions in three rows above and below the waveguide are randomly perturbed with a standard deviation varying between 0 and 6% of the lattice parameter. The Anderson-localized modes are investigated by recording QD photoluminescence spectra under high excitation power where the feeding from multiple QDs makes Anderson-localized modes appear as sharp spectral resonances (Fig. 1C). The observation of spectrally separated random resonances is a signature of Anderson localization of light (*14*), while the detailed statistics of the intensity fluctuations unambiguously verifies localization even in the presence of absorption (*21*). Figure 1D shows the intensity distribution from spectra recorded at different spatial and spectral positions allowing to average over different realizations of disorder. Clear deviations from the Rayleigh distribution predicted for non-localized waves are observed. Light is localized if the variance of the normalized intensity fluctuations exceeds the critical value of 7/3 (*21*), and we extract a variance of 5.3, which clearly proves Anderson localization.

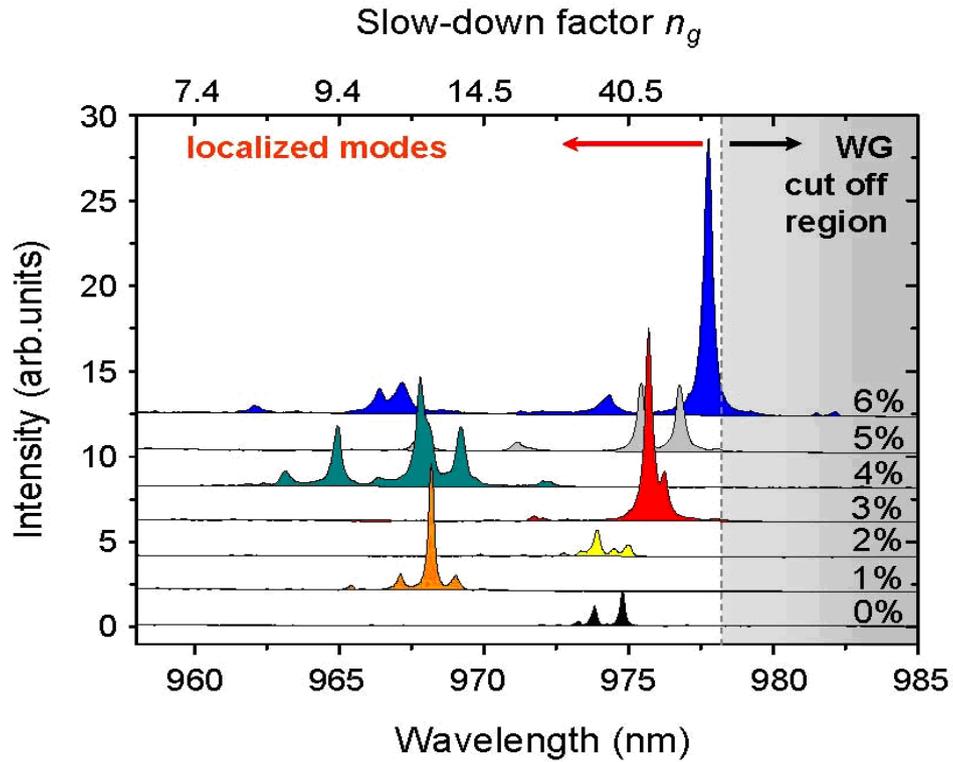

**Fig.2:** Spectral signature of Anderson-localized modes. Photoluminescence spectra collected as in Fig. 1C, for various degrees of engineered disorder. Each spectrum has been collected with the excitation and collection microscope objective at a fixed position on the waveguide and has been vertically shifted for visual clarity. The gray area highlights the calculated waveguide (WG) cut-off region assuming a refractive index of GaAs of 3.44. $n_g = \frac{c}{\partial \omega / \partial k}$ is the calculated group velocity slow-down factor for an ideal structure without disorder, where $c$ is the vacuum light speed, $\omega$ the frequency, and $k$ the wave number.

Examples of Anderson-localized modes are shown in Fig. 2 as peaks appearing at random spectral positions, although limited to the slow-light regime of the photonic crystal waveguide. The latter property is due to the strongly dispersive behaviour of the localization length that is considerably shortened in the slow-light regime. The spectral range of Anderson-localized modes is tuned by controlling the amount of disorder. We note that even in samples without engineered disorder intrinsic and thus unavoidable imperfections, such as surface roughness, are sufficient to localize light (*22*).

The important cavity figures-of-merit are the mode volume $V$ and the $Q$-factor. Decreasing $V$ leads to an enhancement of the electromagnetic field and thus improves light-matter coupling. The $Q$-factor is proportional to the cavity storage time of a photon that needs to be increased for cavity QED applications. High $Q$-factors ranging between 3000 and 10000 are obtained for different degrees of disorder (Fig. 2) and are comparable to state-of-the-art values obtained for traditional photonic crystal nanocavities containing QDs (*7*). Anderson-localized cavities thus offer a fundamentally new route to cavity QED that is inherently robust to fabrication imperfections as opposed to traditional cavities (*9*).

Pumping the sample at low excitation power allows resolving single QD lines, thereby entering the realm of cavity QED. Figure 3A shows an example of a photoluminescence spectrum displaying single QD peaks and Anderson-localized cavities. QDs and cavity peaks can be easily distinguished from their different temperature dependences (Fig. 3B) that also enable the spectral tuning of single QDs into resonance with an Anderson-localized cavity. Fig. 3C displays the crossing between a QD and an Anderson-localized cavity demonstrating that the cavity-QD system is in the Purcell regime where the cavity promotion of vacuum fluctuations enhances the QD decay rate (*6*).

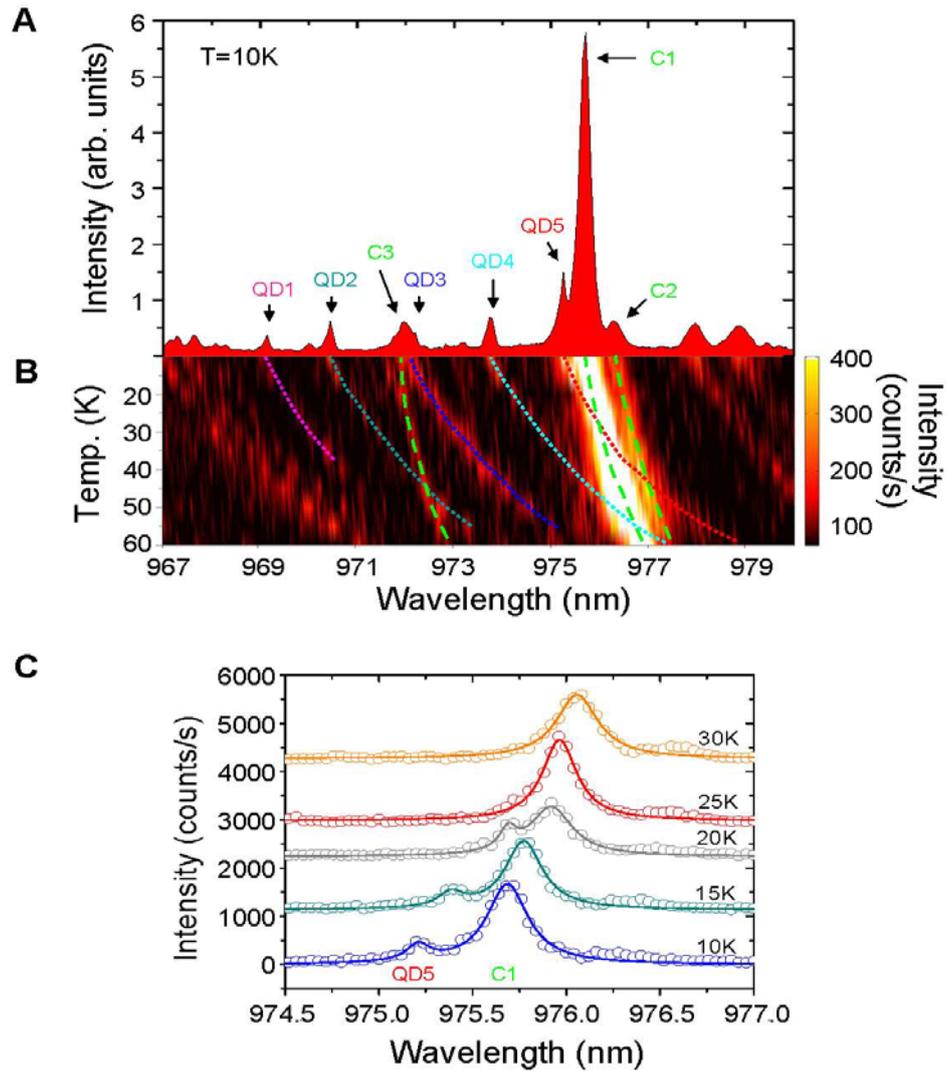

**Fig.3:** Temperature tuning of single QDs into resonance with Anderson-localized cavities. **(A)** Low-power photoluminescence spectrum of a sample with 3% disorder at 10K. **(B)** Photoluminescence spectra collected when varying the sample temperature in steps of 5 K. The dotted (dashed) lines are guides to the eye of the wavelength displacement of selected QD emission lines (localized modes). **(C)** Zoom-in on the spectra displaying the QD-cavity crossing. The spectra are fitted to two Lorentzians (solid lines) representing the QD and the cavity peak.

The Purcell enhancement is studied by means of time-resolved photoluminescence spectroscopy: a QD is repeatedly excited with a short optical pulse and the emission time is measured. Collecting many single-photon events allows recording a decay curve representing a histogram of detection events versus time. Two examples of decay curves for the QD tuned on- and off-resonance with an Anderson-localized cavity are presented in Fig. 4A. Off resonance the QD decay rate is inhibited due to the two-dimensional photonic band gap leading to an emission rate of 0.5 $ns^{-1}$. A pronounced enhancement of a factor of 15 is observed on resonance where a fast decay rate of 7.9 $ns^{-1}$ is extracted. An important figure-of-merit for, e.g., single-photon sources or nanolasers is the *β*-factor, expressing the fraction of photons emitted into a cavity mode. By comparing the emission rates on and off resonance we extract *β* = 94%, which represents a lower bound since even for large detuning residual coupling to the waveguide can persist. The high *β*-factor competes with results obtained on standard photonic crystal nanocavities with carefully optimized cavity design and QD density (*7*). Our results clearly demonstrate that distributed photonic disorder provides a very powerful way of enhancing the interaction between light and matter enabling cavity QED.

The decay rates of two individual QDs tuned across an Anderson-localized cavity are plotted in Fig. 4B. Different enhancement factors (15 and 9 at T = 25 and 55K, respectively) are observed on resonance due to the different positions and dipole orientations of the QDs that influence their coupling to the cavity mode. Note that the presence of an additional Anderson-localized cavity gives rise to the asymmetric detuning dependence of the decay rate. Assuming a perfect spatial match between the QD and the cavity mode, we can extract an upper bound on the mode volume of the Anderson-localized cavity of $V \sim 1$ µm$^3$ from the observed rate on resonance. By estimating the extension of the localized modes in the two directions orthogonal to the waveguide (*23*), we derive a cavity length of 25 µm for cavity C1. Establishing the fundamental lower limit of the cavity length relates to the very fundamental question of determining the localization length of Anderson-localized modes. It is predicted that the localization length can be reduced below the wavelength of light and it was even suggested that no fundamental lower boundary exists *(24)*. Consequently, engineered disorder might pave a way to sub-wavelength confinement of light in dielectric structures.

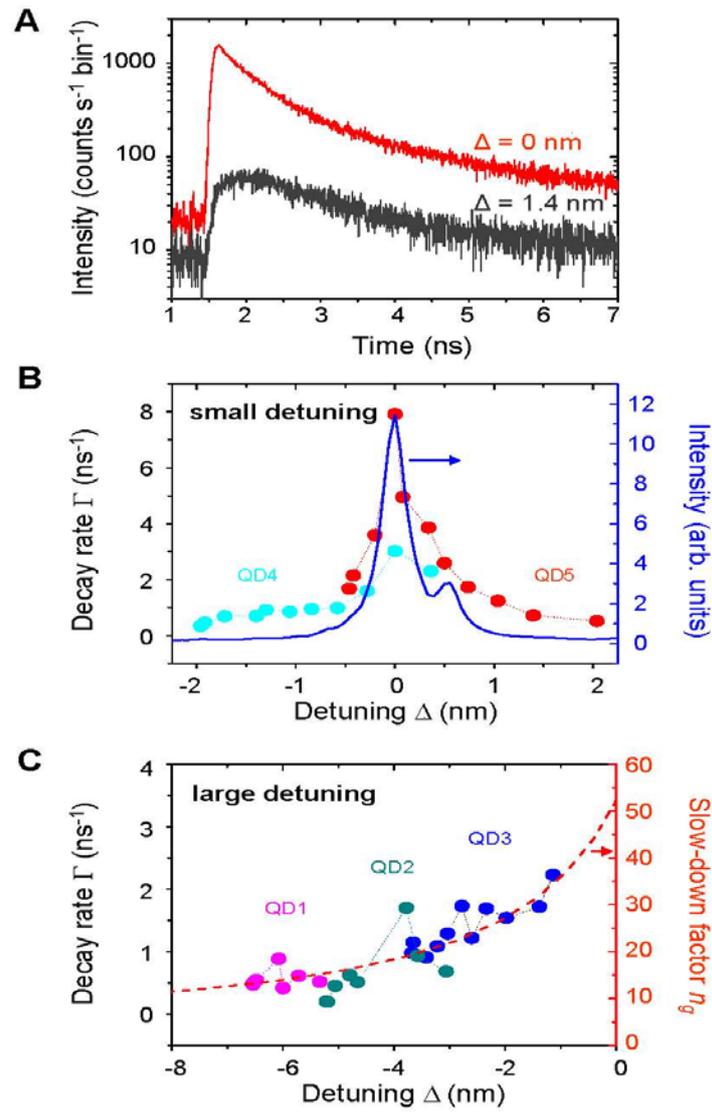

**Fig.4:** Detuning dependence of single QD decay rates. **(A)** Decay curves of QD5 for two values of detuning Δ relative to the localized mode C1. **(B)** Decay rates of QD4 and QD5 versus detuning and cavity emission spectrum. **(C)** Decay rates of QD1-QD3 versus detuning. The dashed line is the calculated slow-down factor for the unperturbed photonic crystal waveguide. The enhancement at Δ = -4 nm stems from the coupling of QD2 to a weak Anderson-localized cavity mode (C3 in Fig. 3A).

Figure 4B shows that Purcell enhancement is observed mainly within the cavity linewidth, which is opposed to the surprisingly far-reaching coupling reported for standard photonic crystal cavities under non-resonant excitation (*7*). Consequently, the extracted QD decay rates are sensitive probes of the local photonic environment of disordered photonic crystal waveguides. Photon emission in disordered photonic structures was predicted to lead to a new class of infinite-range correlations manifested as fluctuations in the decay rate of embedded emitters (*25*). Thus, the Purcell enhancement stems from the local enhancement of the photonic density of states in the Anderson-localized regime that promotes spontaneous emission of photons.

QDs detuned from Anderson-localized cavities may couple to the slowly propagating mode of the photonic crystal waveguide. In this case, the QD decay rate is expected to scale proportional to the group velocity slow-down factor $n_g$ (*26*). This behaviour is observed for three different QDs at large detunings $\Delta$ from the dominating Anderson-localized cavity mode (Fig. 4C), i.e. here the radiative coupling is well described by the local photonic density of states of the unperturbed photonic crystal waveguide. This interesting co-existence of ordered and disordered properties is due to the fact that relatively few periods of the photonic crystal lattice are required to build up the local environment determining the QD decay rate. Thus, the length scale on which the local photonic density of states builds up is mostly shorter than the localization length, which is the reason for the success of photonic crystals despite ubiquitous disorder for, e.g., nanocavities (*9*), single-photon sources (*15*), or spontaneous emission control (*27*).

Our experiments demonstrate that disorder is an efficient resource for confining light in nanophotonic structures opening a new avenue to all-solid-state cavity QED exploiting disorder as a resource rather than a nuisance. Exploring disorder to enhance light-matter interaction and establishing the ultimate boundaries for this new technology provide exciting research challenges for the future, of relevance not only to QED but also to other research fields relying on enhanced light-matter interaction, such as energy harvesting or bio-sensing (*28*). Coupling several cavities is a potential way of scaling cavity QED for quantum information technology and currently represents one of the major challenges for engineered nanocavities. Controlled disorder might offer an interesting route to coherently couple cavities using so-called necklace states that are naturally occurring coupled Anderson-localized modes (*29,30*).

31. We gratefully acknowledge T. Schlereth and S. Höfling for quantum dot growth, J.M. Hvam for discussions, and the Council for Independent Research (Technology and Production Sciences and Natural Sciences) and the Villum Kann Rasmussen Foundation for financial support.

## Materials and Methods

**Fabrication of the photonic crystal structures.** Photonic crystal waveguide membranes in GaAs are fabricated by electron-beam lithography together with dry- and wet-etching. The waveguides are obtained by leaving out a row of holes in the triangular lattice. A controllable degree of disorder is included by varying the position of the holes in the three lines above and below the waveguide using a Gaussian random number generator algorithm (Box-Muller). The amount of induced disorder is evaluated as the root-mean-square of the deviations with respect to the unperturbed hole position divided by the photonic crystal lattice constant ($a$). The samples are $L = 100$ μm long photonic crystal waveguides with lattice constant $a = 260$ nm, hole radius $r = 78$ nm, and membrane thickness 150 nm. A single layer of self-assembled InAs QDs with a density of ~ 80 μm$^{-2}$ is contained in the centre of the membrane.

**Photoluminescence experiments.** The sample is positioned in a Helium flow cryostat and excited with a pulsed Ti:Sapphire laser at 850 nm (pulse length: 2 ps, repetition rate: 75 MHz). A microscope objective (NA = 0.65) is used for the excitation and collection of the emitted light and a CCD spectrograph (spectral resolution: 0.15 nm) and an avalanche photo-diode (temporal resolution: 50 ps) are used for detection. The excitation spot has a measured diameter of 1.4 μm. For further details on the experimental setup, see (*S1*). In the low power excitation experiments, the pump power density is below saturation of the single QDs (~ 20 W/cm$^2$), while in the high power excitation experiments, pump power densities far above saturation are used (~ 2 kW/cm$^2$). To extract the linewidth of the cavity peaks the spectra are deconvoluted with the measured instrument response function. The quantum dot decay rates are extracted by fitting the measured decay curves with several exponential functions. The fastest component extracted from the fit corresponds to the rate of the quantum dot, which is most efficiently coupled to the cavity mode. The slower decay components are due to residual quantum dot recombination processes and quantum dot fine structure (e.g. dark excitons).

**Proving Anderson localization of light.** The proof of Anderson localization of light is obtained from the statistics of the intensity fluctuations of the light scattered vertically out of the photonic crystal waveguides. Photoluminescence spectra are recorded while scanning the microscope objective used for the excitation and collection along the waveguide, see Fig.1c. The quantum dots are pumped with a power density far above saturation to assure homogeneous excitation of the waveguide modes. We collect the intensity of the light leakage $I_j$ at each spatial position $j$ and average over the wavelength range 968-981 nm where the localization length is constant. This procedure is repeated for all the spatial positions with a spatial binning size of 1μm. From these data the probability distribution of the intensity fluctuations $P(I/\langle I \rangle)$ with $I = \sum_j I_j$ can be obtained, see Fig. 1D. Furthermore, the variance $\text{var}(I/\langle I \rangle)$ is extracted, where the average is over realizations of configurations of disorder. The criterion for Anderson localization is: $\text{var}(I/\langle I \rangle) \geq 7/3$, which holds also in the presence of absorptive loss (*S2*). Note that although material absorption of GaAs is negligible at the wavelength of the experiment, several processes in the photonic crystal waveguides act as loss including scattering leakage out of the membrane, residual absorption and scattering from the quantum dots, and the potential formation of absorptive surface states near the GaAs/air interfaces of the membrane.

**Estimating the mode volume.** From the measurement of the modification of the decay rate of a single quantum dot, we can extract the Purcell enhancement ($F_P$) as the emission rate on resonance with the Anderson-localized cavity relative to the rate of emission in free space (*S3*):

$$F_P(\omega) = \frac{3Q(\lambda/n)^3}{4\pi^2 V} \frac{|\vec{d} \cdot \vec{f}(\vec{r}_q)|^2}{|\vec{d}|^2} \frac{1}{1 + 4Q^2(\omega_q/\omega - 1)^2} \quad (S1)$$

where $Q$, $V$, $\omega$, $\lambda$ are the quality factor, mode volume, resonance frequency and resonance wavelength of the Anderson-localized cavity, respectively. $\omega_q$ and $\vec{d}$ denote the emission frequency and transition dipole moment of the quantum dot, while $n = 3.44$ is the refractive index of surrounding GaAs substrate and $\vec{f}(\vec{r}_q)$ is the electric field distribution of the cavity mode evaluated at the quantum dot location $\vec{r}_q$. The Purcell factor for QD5 coupled to the Anderson-localized cavity C1 (see Fig. 4B) is $F_P = 7.2$ since the spontaneous emission rate of a quantum dot in a homogeneous medium is measured to be 1.1 ns$^{-1}$. Using the measured $Q = 4200$ and assuming perfect spatial match between the quantum dot transition dipole moment and the spatial mode profile of the cavity, we experimentally determine an upper bound on the mode volume of the Anderson-localized cavity mode C1 of $V = 1$ μm$^3$.

Based on the experimental value of the mode volume we can extract the spatial extent of the Anderson-localized cavity mode along the waveguide. The mode volume is defined as: $V = \frac{1}{n^2} \int_{\vec{r}} n(\vec{r})^2 |\vec{f}(\vec{r})| d^3\vec{r}$. An in-plane extension of the mode perpendicular to the waveguide of 1.3±0.1 μm is measured from the photoluminescence data collected while scanning the microscope objective across the waveguide. The vertical extension of the mode out of the membrane is calculated numerically using ab-initio plane wave calculations considering an ideal photonic crystal structure with no engineered disorder. From these considerations a spatial extent of 25 μm along the direction of the photonic crystal waveguide is obtained for the Anderson-localized cavity C1.

**Extracting the β-factor.** The figure-of-merit for the coupling of the emitter to the Anderson-localized cavity is the β-factor. It is defined as the ratio between the decay rate of a quantum dot into the localized mode ($\Gamma$) and the total decay rate ($\Gamma_{tot}$):

$$\beta = \frac{\Gamma}{\Gamma + \gamma_{rad} + \gamma_{nr}} = \frac{\Gamma_{on} - \Gamma_{off}}{\Gamma_{on}} \tag{S2}$$

where $\gamma_{rad}$ is the rate of a quantum dot coupling to radiation modes and $\gamma_{nr}$ its intrinsic non-radiative decay rate. $\Gamma_{on}$ ($\Gamma_{off}$) represents the experimentally extracted decay rate of a single quantum dot on (off) resonance with the localized cavity mode.